# THE DYNAMICAL SURVIVAL OF SMALL-SCALE SUBSTRUCTURE IN RELAXED GALAXY CLUSTERS

Guillermo González-Casado[1], Gary A. Mamon[2,3] and Eduard Salvador-Solé[4]

## ABSTRACT

We consider the dynamical evolution of small-scale substructure in clusters within two extreme alternate scenarios for their possible origin: 1) the accretion of groups (or small clusters) on quasi-radial orbits, and 2) the merger of clusters of similar masses, followed by the decoupling of their dense cores. Using simple analytical arguments and checking with numerical computations, we show that objects are destroyed by the tidal field of the global cluster potential if their mean density is small compared to the mean cluster density within the radius of closest approach of the group or detached core. Accreted groups and small clusters are thus tidally disrupted in one cluster crossing. Since the cores of clusters are much denser than groups, they are considerably more robust to tides, but the least massive are destroyed or severely stripped by tides, while the others are brought to the cluster center by dynamical friction (and subsequently merge) in less than one orbit. The longest lived substructures are detached cores, roughly ten times less massive than the cluster, starting in near-circular orbits beyond $1\,h^{-1}$ Mpc from the cluster center.

*Subject headings:* galaxies: clustering — galaxies: kinematics and dynamics — cosmology: theory

## 1. INTRODUCTION

Clusters of galaxies are often complex structures harboring two different kinds of probably related substructure. First, *large-scale substructure*, appearing as a bimodal grand design, has been reported in roughly 30% of clusters from the study of projected galaxy positions, redshifts, and X-ray emission maps (*e.g.* Dressler & Shectman 1988b; Jones & Forman 1992; Escalera *et al.* 1994). Second, *small-scale substructure*, in the form of density enhancements around luminous galaxies seems also present in the central regions of some otherwise apparently relaxed clusters. Such small-scale substructure is apparent in the Coma cluster (Fitchett & Webster 1987; Mellier *et al.* 1988; White, Briel, & Henry 1993), and often inferred from gravitational optics analyses of giant arcs (Miralda-Escudé 1993; Kneib *et al.* 1993). A fraction of $\simeq 50\%$ of large-scale monomodal clusters could harbor small-scale substructure according to the statistical study by Salvador-Solé, Sanromà, & González-Casado 1993, hereafter SSG) of Dressler & Shectman's (1988a) sample of clusters, confirmed by the analysis of correlation among galaxy positions within those clusters (Salvador-Solé, González-Casado, & Solanes 1993, hereafter SGS), and from wavelet analysis (Escalera *et al.* 1994).

The presence of large-scale bimodal substructure strongly suggests that clusters are caught in the process of merging. Thus, the amount of clusters harboring large-scale substructure is closely related with the rate of formation of massive cosmic structures. Following this idea, three recent studies (Richstone, Loeb, & Turner 1992; Lacey & Cole 1993; Kauffmann & White 1993) have used the observed fraction of large-scale substructure in clusters to constrain the value of $\Omega_0$ (basically guessing the survival times of substructures), and while the first two concluded that $\Omega_0 > 0.5$, Kauffmann & White (1993) remarked that the survival time of substructure is too uncertain to conclude on $\Omega_0$. In this *Letter*, the dynamical survival of observed small-scale substructure in clusters is assessed with attention to its possible origin.

In the framework of hierarchical clustering, small-scale substructure could appear in apparently relaxed clusters in two ways. First, this substructure could simply be a group or small cluster *accreting* onto the cluster. A more subtle idea (González-Casado, Solanes & Salvador-Solé 1993, hereafter GSS) is to have two clusters of similar mass *merge* together, and in that process, their central cores detach from their main body. This type of dynamical evolution is seen in simulations of galaxies merging together: their extended dark halos merge before their more bound visible components. (*e.g.* Barnes 1992). Indeed, the violent relaxation induced by the merger generates rapid variations in the global potential, subjecting different lagrangian elements to rapidly varying tides. Although a detailed study of tidal processes during merging is well beyond the scope of this *Letter*, it is safe to assume that, if the initial clusters have cuspy cores, these are indeed likely to survive the tidal processes during the violent relaxation phase occurring during the merger. Both accretion and merging scenarios are studied below.

## 2. DYNAMICS OF IMMERSED SUBSTRUCTURE

Groups or small clusters accreting onto larger clusters as well as those detached cores that follow elongated orbits within a cluster (both subsystems are hereafter called *clumps*) are subject to a tidal shock from the global potential of the cluster as they pass by its center.

The tidal acceleration of a galaxy in the clump can be written as

$$\mathbf{a}_{\rm tid} = -\frac{GM(S)}{S^3}\mathbf{S} + \frac{GM(R)}{R^3}\mathbf{R} = \frac{GM(R_p)}{R_p^2}\mathbf{q}, \quad (1)$$

where $\mathbf{R}$ and $\mathbf{S}$ are the positions of the clump center and one galaxy within the clump in the frame of the cluster center, respectively, $R_p$ is the orbit pericenter, and where

$$\mathbf{q} = -\frac{f(S)}{(S^2/R_p^2)}\hat{\mathbf{S}} + \frac{f(R)}{(R^2/R_p^2)}\hat{\mathbf{R}},$$

[1] Departament de Matemàtica Aplicada II, Universitat Politècnica de Catalunya, Pau Gargallo 5, E–08028 Barcelona, Spain.
[2] Institut d'Astrophysique, 98 bis Blvd Arago, F–75014 Paris, France.
[3] DAEC, Observatoire de Paris-Meudon, F–92195 Meudon, France.
[4] Departament d'Astronomia i Meteorologia, Universitat de Barcelona Avda. Diagonal 647, E–08028 Barcelona, Spain.

with $f(R) = M(R)/M(R_p)$, while $\hat{\mathbf{S}}$ and $\hat{\mathbf{R}}$ are unit vectors along $\mathbf{S}$ and $\mathbf{R}$, respectively. Integrating over time, assuming that clumps remain stationary, and working in cartesian coordinates, one can obtain the velocity increment of galaxies in a clump. Then, averaging these increments over the galaxy distribution of the clump and assuming an impulsive approximation (Spitzer 1958; see also Mamon 1987) leads to a gain in internal energy

$$\Delta U = \frac{1}{2} m_s \langle (\Delta v)^2 \rangle = \frac{m_s}{2} \left[ \frac{GM(R_p)r_s}{R_p^2 V_p} \right]^2 \langle (\Delta w)^2 \rangle \;,$$

where $m_s$ and $r_s$ are the mass and radius of the clump, respectively, $V_p$ is the clump's velocity at pericenter, and $\Delta w$ is a dimensionless velocity increment:

$$\Delta w = \int_0^T \mathbf{q} \, \frac{dt}{(r_s/V_p)} \;, \qquad (2)$$

with $T$ the time spent by the clump between two consecutive apocenters. Writing the clump's internal energy as $U = -\gamma G m_s^2/r_s$, one finally obtains

$$\frac{\Delta U}{U} = -\frac{\langle (\Delta w)^2 \rangle}{2\gamma} \left[ \frac{V_{\rm circ}(R_p)}{V_p} \right]^2 \frac{\bar{\rho}_{\rm cl}(R_p)}{\bar{\rho}_s(r_s)} \;, \qquad (3)$$

where $\bar{\rho}_{\rm cl}$ and $\bar{\rho}_s$ are the respective mean densities of the cluster and substructure, while $V_{\rm circ}$ is the circular velocity of the cluster. Hence, *the relative internal energy gain scales as the ratio of the mean densities of cluster within pericenter to substructure* (see also Mamon 1992). Equation (3) implies that *groups or small clusters that penetrate deeply into the cluster will suffer tides more severely than detached cuspy cores*, since the mean density of a cluster core is much higher than the mean density of a whole group or cluster. Only groups as compact as Hickson's (1982) compact groups may be robust to tidal destruction, since they appear as dense as the cores of rich clusters.

We integrate numerically the orbit of the clump in the cluster potential, including the loss of orbital energy of the clump by dynamical friction against the dark matter particles of the cluster. The force of dynamical friction is

$$\mathbf{F_{df}} = -2\pi G^2 m_s^2 \ln(1+\Lambda^2) \rho_{\rm cl}(R) \frac{[{\rm erf}(x_0) - x_0 {\rm erf}'(x_0)]}{v^2} \hat{\mathbf{v}} \;,$$

where $x_0 = 2^{-1/2} v/\sigma_{\rm cl}$, $\rho_{\rm cl}(R)$ and $\sigma_{\rm cl}(R)$ are the mass density and 1D velocity dispersion of the cluster at a distance $R$ from its center, and $\Lambda = p_{\max}/p_{\min}$, is the ratio between the maximum and minimum impact parameters for which collisions between the subsystem and the light background particles are efficient. Here, we assume either 1) $p_{\max} = R$ and $p_{\min} = Gm_s/v^2$ for clusters and clumps with a singular isothermal density profile (Tremaine 1976; Lacey & Cole 1993); or 2) $p_{\max} = \max(R_c, R)$ and $p_{\min} = r_c$ for clusters and clumps with a flat core of radius equal to $R_c$ and $r_c$, respectively (Tremaine, Ostriker & Spitzer 1975).

The tidal acceleration (eq. [1]) is evaluated along the orbit, from first to second apocenter. Velocity increments are calculated by numerical integration of equation (2) and the relative internal energy gain is computed through equation (3). A subsystem with $|\Delta U/U| > 1$ after one orbit is considered to be disrupted within a semi-orbital time, $T/2$. Indeed, clumps destroyed by tidal forces will dissolve in a timescale shorter than their internal semi-orbital time, which itself is shorter than $T/2$. Clumps surviving cluster tides and very massive clumps will have their orbits decay by dynamical friction towards the cluster center. In these cases, the survival time is set equal to the orbital decay time, defined as the time for the clump to definitely settle within the cluster core radius.

## 3. ACCRETED GROUPS

From spherical cosmological infall, a group accreted on a radial orbit by a cluster and reaching its pericenter at present time, $t_0$, had an apocenter that was the cluster turnaround radius at time $t_0/2$ (Gunn & Gott 1972): $R_a = \left(2GM_{\rm cl} t_0^2/\pi^2\right)^{1/3}$, where $M_{\rm cl}$ is the present cluster mass within the radius, $R_1 = R_a/2$, of a circular orbit with period equal to $t_0$. This roughly approximates the size of the region which may be relaxed in the cluster (White 1992). The cluster crossing time is defined as $t_{\rm cr} = R_1/V_{\rm circ}(R_1) = t_0/(2\pi)$.

In this accretion scenario, the cluster is assumed to have a cuspy density profile, $\rho_{\rm cl} \sim R^{-2}$, consistent with gravitational optics analyses (*e.g.* Hammer 1991). We adopt the same cuspy density profile for infalling groups or small clusters, and their radii are inferred in the same way as the cluster radius. This contradicts the reported absence of central concentration in the density profile of small groups (Walke & Mamon 1989), therefore, the tidal disruption times derived for groups will be upper limits, since we are overestimating their central concentration and, therefore, their ability to survive tides.

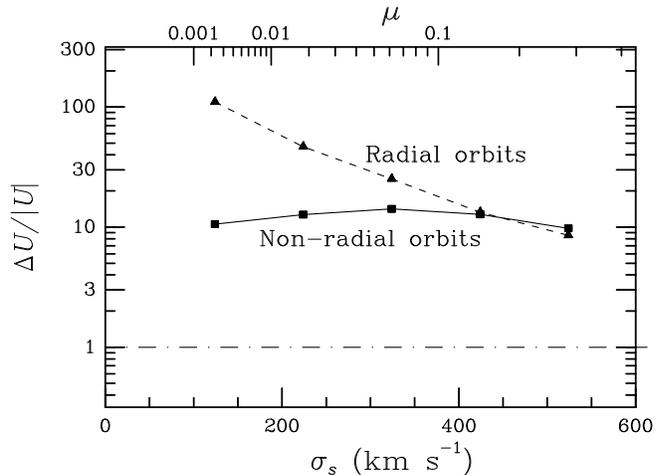

FIG. 1.— Relative internal energy increment after a pericentric passage of accreted groups of internal velocity dispersion $\sigma_s$, with apocenter $R_a = 2.2 h^{-1}$ Mpc. The initial apocentric velocities are $250\,{\rm km\,s^{-1}}$ (*squares*) and $50\,{\rm km\,s^{-1}}$ (*triangles*), corresponding to specific angular momenta $S = (J/J_{\rm circ}[E])^2 = 0.2$ and $0.01$, respectively. The mass fraction, $\mu = m_s/M_{\rm cl}$, is given in the top axis. The horizontal *dashed-dotted line* is for critical disruption.

Figure 1 shows the relative internal energy gains of accreted groups and small clusters of different velocity dispersions (or mass $m_s = \mu M_{\rm cl}$) and initial circularity parameters (defined as the squared ratio of the initial angular momentum to the angular momentum of the circular orbit of same energy, Merritt 1985). We adopt a 1D cluster velocity dispersion of $\sigma_{\rm cl} = 760\,{\rm km\,s^{-1}}$, yielding $R_1 = 2^{-1/2}\pi^{-1}\sigma_{\rm cl}t_0 = 1.1\,h^{-1}$ Mpc and $M_{\rm cl} = 2^{1/2}(\pi G)^{-1}\sigma_{\rm cl}^3 t_0 = 3.1 \times 10^{14}\,h^{-1}\,M_\odot$, for $\Omega_0 = 1$. Figure 1 clearly shows that *infalling groups and small clusters are tidally disrupted at first passage through the core of a rich cluster, regardless of their initial orbital parameters*, since they gain considerable more energy than their binding energy. This result is independent of the values assumed for $\Omega_0$ and $\sigma_{\rm cl} \gtrsim 400\,{\rm km\,s^{-1}}$. A similar result has recently been reported (Capelato & Mazure 1995) through use of $N$-body simulations (with a non-cuspy cluster potential). Groups accreted on nearly radial orbits are more easily disrupted because their pericenters are closer to the cluster center.

## 4. DETACHED CORES OF MERGED CLUSTERS

We now evaluate numerically the survival time of a clump, initially immersed in a cluster, similarly as for accreted groups, except that: 1) the cluster density profile is flat at the center, since cores have detached, and 2) the initial distance to the cluster center and the initial specific orbital angular momentum are free parameters, since, after violent relaxation, the clumps are left orbiting with random position and velocity direction. Since violent relaxation leaves a phase space distribution independent of mass (Lynden-Bell 1967), the typical initial velocity $V_i$ of clumps will be the 3D velocity dispersion of galaxies in the cluster.

We adopt, for the cluster, a Modified Hubble number density profile $n(R) = n_0/[1 + (R/R_c)^2]^{3/2}$ and constant galaxy velocity dispersion $\sigma_g$, both consistent with observations. Assuming isotropic hydrostatic equilibrium, the gravitational potential of the cluster mass distribution is then $\Phi(R) = (3/2)\sigma_g^2 \ln[1 + (R/R_c)^2]$. The resulting (essentially dark matter) background mass density and 1D velocity dispersion are

$$\rho_{\rm cl}(R) = \frac{3\sigma_g^2}{4\pi G R_c^2} \frac{3 + X^2}{(1 + X^2)^2} \qquad \sigma_{\rm cl}^2(R) = \sigma_g^2 \frac{3(2 + X^2)}{2(3 + X^2)},$$

where $X = R/R_c$, $R_c = 0.25\,h^{-1}$ Mpc is the cluster core radius, and where $\sigma_{\rm cl}$ is obtained by solving the isotropic hydrostatic equation for the background mass. The same model has been assumed for the detached core, but with a smaller core radius, $r_c = 0.06\,h^{-1}$ Mpc, following the optical properties of clumps obtained by GSS from a simple clustering model that reproduces the statistical properties of the observed small-scale substructure in clusters. According to SGS and GSS, the small-scale substructure in clusters occurs in clumps smaller than $0.3\,h^{-1}$ Mpc, and seems to harbor typically about 10% of the galaxies in the central $1\,h^{-1}$ Mpc region of clusters.

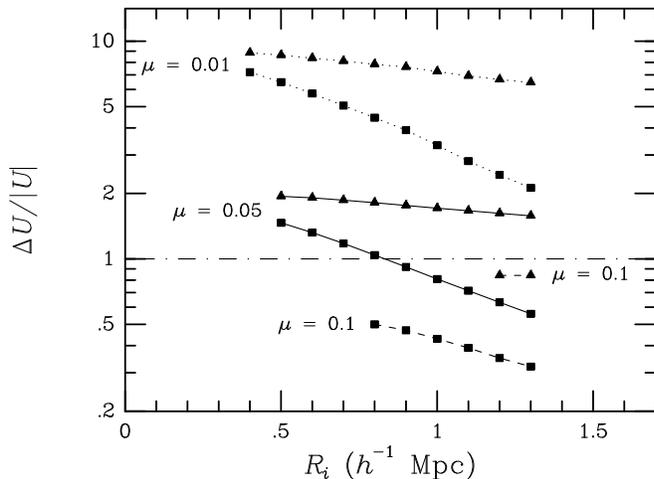

FIG. 2.— Relative internal energy increment of detached cores of size $r_s = 0.2\,h^{-1}$ Mpc *vs.* initial distance to the cluster center, $R_i$, for different mass fractions $\mu = m_s/M_{\rm cl}$. *Triangles* represent radial orbits, and *squares* are for orbits with specific angular momentum $S = 0.25$. The horizontal *dashed-dotted line* is for critical tidal disruption. Only cases with at least one secondary apocenter are shown.

Figure 2 shows the relative gain in internal energy after a pericenter passage for detached cores on elongated orbits, with mass $m_s = \mu M_{\rm cl}$, for $r_s = 0.2\,h^{-1}$ Mpc (we have tried different values of $r_s$). *Massive clumps with $\mu \gtrsim 0.05$ and $r_s \lesssim 0.25\,h^{-1}$ Mpc survive the cluster tidal field, whereas low-mass clumps do not*. For nearly circular orbits, the tide is not impulsive but roughly stationary, and the clump may be tidally limited to some radius. Severely limited clumps cannot be observed as small-scale substructure. The lower limit of the tidal radius is obtained when the clump's orbit is circular, and its internal rotation is phase-locked with the orbit (see Mamon 1987; Allen & Richstone 1988).

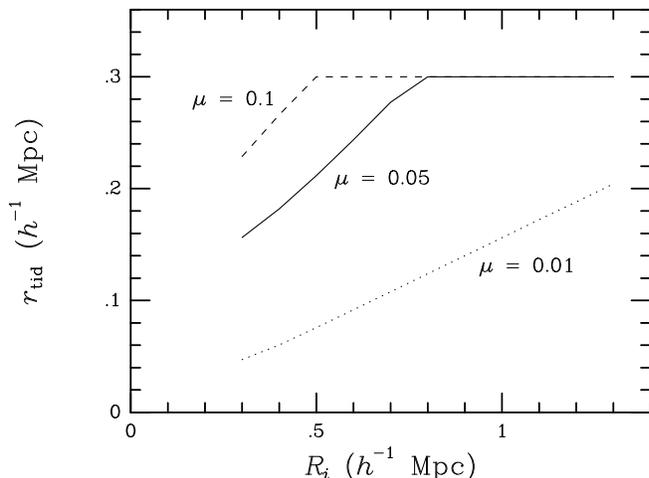

FIG. 3.— Tidal stripping radius of clumps moving in different phase-locked circular orbits of radius $R_i$, for different mass fractions $\mu = m_s/M_{\rm cl}$. The initial clump size is $0.3\,h^{-1}$ Mpc.

Figure 3 shows the tidal radius for such an orbit computed from King's (1962) 'instantaneous' criterion. Tidal stripping is negligible for $\mu \gtrsim 0.05$, except for very bound clumps ($R_i \lesssim 0.6\,h^{-1}$ Mpc). Note that during violent relaxation, tides arising from transient density distributions probably lead to gradients of the gravitational potential similar to those found in the central parts of relaxed clusters. Hence, clumps that do not survive to impulsive tides from the potential of a relaxed cluster would have hardly survived to the previous process of cluster merging. Although the more massive clumps are more robust to tides, they are also more prone to dynamical friction, which causes their orbits to decay towards the cluster core where they disappear as substructures.

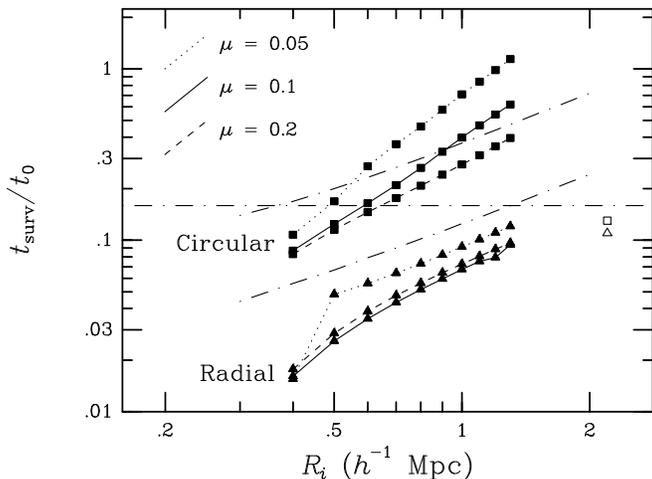

FIG. 4.— Survival time (either to settle within $R < R_c$ or before being disrupted by tides) of detached cores vs. initial distance for radial (filled triangles) and circular orbits (filled squares), and for different mass fractions $\mu = m_s/M_{cl}$. The dashed-dotted curves and horizontal dashed-dotted line denote the clump circular (upper) and radial (lower) unperturbed semi-orbital times and cluster crossing time (defined in § 3), respectively. The open triangle and square are the times spent within the cluster ($R < R_1$), before tidal disruption, by accreted groups, respectively on radial and non-radial orbits (see § 3).

Figure 4 shows the survival time (defined in § 2) for clumps of different mass. We adopt $V_i$ moving inwards and the same velocity dispersion for the cluster galaxies as in § 3, yielding $R_1 = 1.4\,h^{-1}$ Mpc and $M_{cl} = 5.5 \times 10^{14}\,h^{-1}\,M_\odot$ for $\Omega_0 = 1$, but a larger value of $\sigma_g$ or $\Omega_0 < 1$ only cause slight changes in the results. When the clump survives the tides, ($\mu \gtrsim 0.07$) the orbital decay time is shorter than an orbital period, and the clump falls directly into the cluster center. Nevertheless, clumps with $0.05 < \mu < 0.2$ and $R_i \gtrsim 0.6\,h^{-1}$ Mpc have a survival time longer than one cluster crossing time when their initial orbits are not too elongated ($S \gtrsim 0.3$). Furthermore, only clumps with $r_s \lesssim 0.25\,h^{-1}$ Mpc survive the cluster tides, which agrees with the upper limit to the sizes of detached cores derived from observation (SGS; GSS).

## 5. DISCUSSION

Small-scale substructure could be made of groups or small clusters that have been accreted since less than one cluster crossing time, and are about to be tidally disrupted. In this case, a high rate of accretion would be required to reproduce the observed high fraction (roughly 50%, § 1) of clusters with small-scale substructure.

On the other hand, small-scale substructure may be caused by detached cores that have survived the violent relaxation after a merger of two similar sized clusters. According to the survival times computed in § 4, the merger event could have taken place during the last 3 or 4 cluster crossing times for the longest-lived clumps, which are obtained by a fine-tuned balance between tides and dynamical friction for $\mu \simeq 0.1$–$0.15$ and initial circular orbits. The probability that a detached core with $\mu = 0.1$ has $t_{\rm surv} \geq t_{\rm cr}$ is roughly 40%, if we assume that detached cores after violent relaxation are distributed with the same density profile as galaxies in the cluster (SGS), and with random initial orbit elongations. Therefore, the observed high fraction of detection of small-scale substructure may be reproduced, provided a high rate of cluster mergers.

Nonetheless, the true origin of the observed small-scale substructure depends not only on the dynamical survival times, but also on the relative importance of the rates of accretion of small clusters versus the merging with clusters of similar mass, both of which depend on the power spectrum of the primordial density fluctuations and on $\Omega_0$. In a forthcoming paper, we will compare the survival timescales derived here with the observed frequency of small-scale substructure in apparently relaxed clusters to place important cosmological constraints.

We thank the referee, Eliot Malumuth, for useful comments. This work was supported in part by CICYT projects PB 90–0448 and PB 92–0794.